**Spin Depolarization in Quantum Wires Polarized Spontaneously in a Zero Magnetic Field**

N.T. Bagraev[1], V.K. Ivanov[2], L.E. Klyachkin[1], and I.A. Shelykh[2,3]

[1]*A.F.Ioffe Physicotechnical Institute, 194021, St. Petersburg, Russia*

[2]*St. Petersburg State Polytechnical University, 195251, St. Petersburg, Russia*

[3]*LASMEA UMR 6602, Université Blaise Pascal,24, av. des Landais, 63177, Aubiere, France*

**Abstract**

The conditions for a spontaneous spin polarization in a quantum wire positioned in a zero magnetic field are analyzed under weak population of one-dimensional subbands that gives rise to the efficient quenching of the kinetic energy by the exchange energy of carriers. The critical linear concentration of carriers above which the quasi one-dimensional gas undergoes a complete spin depolarization is determined by the Hartree-Fock approximation. The dependence of the critical linear concentration on the carrier's concentration is defined to reveal the interplay of the spin depolarization with the evolution of the "$0.7 \cdot (2e^2/h)$" feature in the quantum conductance staircase from the $e^2/h$ to $3/2 \cdot (e^2/h)$ values. This dependence is used to study the effect of the hole concentration on the "$0.7 \cdot (2e^2/h)$" feature in the quantum conductance staircase of the quantum wire prepared inside the p-type silicon quantum well using the split-gate technique. The 1D channel is demonstrated to be spin-polarized at the linear concentration of holes lower than the critical linear concentration, because the "$0.7 \cdot (2e^2/h)$" feature is close to the value of $0.5 \cdot (2e^2/h)$ that indicates the spin degeneracy lifting for the first step of the quantum conductance staircase. The "$0.7 \cdot (2e^2/h)$" feature is found to take however its normal magnitude when the linear concentration of holes attains the critical value corresponding to the spin depolarization. The variations in the height of the "$0.7 \cdot (2e^2/h)$" feature observed in the hole quantum conductance staircase that is revealed by the p-type silicon quantum wire seem to be related to the evidences of the quantum conductance staircase obtained by varying the concentration of electrons in the 1D channel prepared inside the GaAs-AlGaAs heterojunction [1].

**Spin Depolarization in Quantum Wires Polarized Spontaneously in a Zero Magnetic Field**

N.T. Bagraev[1], V.K. Ivanov[2], L.E. Klyachkin[1], and I.A. Shelykh[2,3]

[1]A.F.Ioffe Physicotechnical Institute, 194021, St. Petersburg, Russia

[2]St. Petersburg State Polytechnical University, 195251, St. Petersburg, Russia

[3]LASMEA UMR 6602, Université Blaise Pascal,24, av. des Landais, 63177, Aubiere, France

**1. Introduction.**

Progress in nanotechnology makes it possible to fabricate low-dimensional semiconductor systems with low density of high-mobility charge carriers, which exhibit ballistic behavior under the condition $k_B T \tau / \hbar > 1$, where $\hbar / k_B T$ is the time of electron-electron interaction and $\tau = m^* \mu / e$ is the transport relaxation time [1-15]. In contrast to the diffusion mode ($k_B T \tau / \hbar < 1$), the role of spin correlations is considerably enhanced in the processes of ballistic transport [2-6]. Among their most dramatic manifestations in the localization and transport processes is the appearance of the "$0.7 \cdot (2e^2 / h)$" feature, which is split off from the first step in the quantum conductance staircase revealed by a one-dimensional (1D) channel [1, 7-9].

The charge transport in such channels that are prepared by the split-gate [10-12] and cleaved edge overgrowth [13] methods is not accompanied with the Joule losses, because their length is less than the mean free path. Therefore, the conductance of the quantum wire that contains a single or several ballistic 1D channels depends only the transmission coefficient $T$ [14,15]; i.e.,

$$G_0 = g_s \frac{e^2}{h} N \cdot T \qquad (1)$$

Where $N$ denotes the number of the highest occupied 1D subband, which is changed by varying the gate voltage, $U_g$. Furthermore, the dependence $G(U_g)$ represents the quantum conductance staircase, because the conductance of a quantum wire is changed by the value of $g_s e^2 / h$ each time when the Fermi level coincides with one of the 1D subbands [11,12]. Spin factor, $g_s$, describes the spin degeneration of the wire mode. The value of $g_s$ is equal to two for non-interacting fermions if the external magnetic field is absent and becomes unity as a result of Zeeman splitting of a quantum staircase in strong magnetic field. The first step of the quantum conductance staircase has been found, however, to split into two parts even in the absence of external magnetic field [1,7-9]. The height of the substep that is dependent on temperature is usually observed to be about 0.7 of the first step value in a zero magnetic field.

Two experimental observations indicate the importance of the spin component for the behavior of this "$0.7 \cdot (2e^2/h)$" feature. First, the electron $g$ - factor was found to increase from 0.4 to 1.3 as the number of occupied 1D subbands decreases [7]. Second, the height of the "$0.7 \cdot (2e^2/h)$" feature attains to a value of $0.5 \cdot (2e^2/h)$ with increasing external magnetic field [7-9]. These results have defined the spontaneous spin polarization of a 1D gas in a zero magnetic field as one of possible mechanisms for the "$0.7 \cdot (2e^2/h)$" feature in spite of the theoretical prediction of a ferromagnetic state instability in ideal 1D systems in the absence of a magnetic field [16].

The studies of the spontaneous spin polarization in quantum wires that have been carried out recently in frameworks of the Kohn-Sham mean-field approximation with the ultra-low linear concentration of charge carriers when the energy of the exchange interaction begins to exceed the kinetic energy in a zero magnetic field allowed to describe qualitatively the quantum conductance staircase of a spin-polarized 1D channel [17-23]. However, the behavior of the "$0.7 \cdot (2e^2/h)$" feature with the onset of spin depolarization that is enhanced by increasing the linear concentration of charge carriers specifically at finite temperatures is still elusive. Here we use the Hartree-Fock approximation to define the critical linear concentration of carriers above which the spin-polarized quasi one-dimensional gas undergoes the spin depolarization that seems to be related to the evolution of the "$0.7 \cdot (2e^2/h)$" feature in the quantum conductance staircase from the $e^2/h$ to $3/2 \cdot (e^2/h)$ values. The energies of the completely spin-polarized and unpolarized states of the quasi one-dimensional gas are analytically calculated and then compared. The completely spin-polarized state is shown to be stable when the linear concentration of carriers is lower than the critical value, above which the quasi one-dimensional gas undergoes the spin depolarization. This result allows to explain the variations in the height of the "$0.7 \cdot (2e^2/h)$" feature that have been found by varying the concentration of electrons in the 1D channel prepared inside the GaAs-AlGaAs heterojunction [1]. Finally, the effect of the concentration of the charge carriers on the "$0.7 \cdot (2e^2/h)$" feature is experimentally verified in the studies of the hole quantum conductance staircase revealed by the quantum wire prepared inside the p-type silicon quantum well using the split-gate technique.

**2. Spontaneous spin polarization in one-dimensional system in a zero magnetic field.**

The 1D systems of fermions are described by the Schrödinger equation $H\Psi = E\Psi$ with $H = H_0^{1D} + H_1$; where $H_0^{1D}$ is the Hamiltonian of non-interacting fermions, which depends on the dimensionality of the system under study, and the term $H_1$ accounts for the interaction of fermions. The form of $H_0^{1D}$ depends on the dimensionality of the system under consideration. Since the motion of charge carriers in 1D systems is quantized in two directions (*x*, *y*), the Hamiltonian for non-interacting particles has the form

$$H_0^{1D} = \sum_{j=1}^{N}\left(\frac{\mathbf{p}_j^2}{2m} + U(x_j, y_j)\right) \tag{2}$$

The corresponding unperturbed wave functions can be written as

$$\psi_{k,m}(\mathbf{r}) = \frac{1}{\sqrt{\Omega_{1D}}} e^{ikz} \varphi_m(\boldsymbol{\rho}) \tag{3}$$

where *m* is the number of the subband of the quantum confinement in the plane $\mathbf{r} = \mathbf{i}x + \mathbf{j}y$, and $\Omega_{1D}$ is the 1D volume that is a quantity with the dimensionality of length.

The interaction operator that is independent of the dimensionality is given by

$$H_1 = \frac{1}{2}\sum_{i \neq j} V(|\mathbf{r}_i - \mathbf{r}_j|) \tag{4}$$

where

$$V(|\mathbf{r}_i - \mathbf{r}_j|) = \frac{e^2}{|\mathbf{r}_i - \mathbf{r}_j|} \tag{5}$$

In the second-quantized representation,

$$H_1 = \frac{1}{2} \sum_{i \neq j} \langle KL|V|MQ \rangle c_K^+ c_L^+ c_Q c_M \tag{6}$$

where each of the subscripts $K$, $L$, $M$, and $Q$ stands for a particle wave vector, the number of the quantum-confinement subband, and the spin.

If the density of non-interacting carriers is sufficiently low so that only the lowest quantum-confinement subband is occupied, the total energy of the electron gas equals its kinetic energy, and the energy density can easily be calculated both for a 1D gas of charge carriers:

$$\varepsilon_{kin} = \sum_{k<k_F} \frac{\hbar^2 k^2}{2m} \tag{7}$$

where $k_F$ is the Fermi wave vector.

Accordingly,

$$\varepsilon_{kin}^{1D} = \sum_{k<k_F} g_s \frac{\hbar^2 k^2}{2m} = \frac{\hbar^2 g_s}{12\pi m} k_F^3 = \frac{\pi^2 \hbar^2 n_{1D}^3}{12 m g_s^2} \tag{8}$$

where the value of $k_F$ is determined from the condition

$$k_F = \frac{\pi}{g_s} n_{1D} \tag{9}$$

Here $n_{1D}$ is the linear concentration of charge carriers in a 1D gas and $g_s$ is the spin factor, which is equal to the number of electrons per unit cell of phase space. For an unpolarized state, $g_s = 2$, whereas for a completely spin-polarized state, $g_s = 1$. The values $1 < g_s < 2$ correspond to the partial spin polarization of charge carriers in a 1D gas.

The spin-polarized state of a 1D gas of non-interacting fermions seems to be unfavorable energetically, because its kinetic energy is always higher than the kinetic energy of an unpolarized state. Therefore the spontaneous spin polarization due to the exchange interaction in quasi one-dimensional systems appears to be unlikely. However, the additional energy term $E_1$ that can be represented by the following infinite sequence of diagrams

$$\mathcal{E}_{core} + \mathcal{E}_{el-el} = \;\diagram_1 + \diagram_2 + \diagram_3 + \diagram_4 + \diagram_5 + ... \qquad (10)$$

has to be taken into account if a 1D gas consists of interacting particles. The exchange diagrams 3 and 5 are easily seen to be dependent substantially on the spin polarization of the system. Indeed, the interaction is independent of the spin, which implies spin conservation at the diagram vertices. Thus, only particles with the same spin may be involved in the processes described by the exchange diagrams, the contribution of which is more significant to the spin-polarized than to unpolarized systems. Since the contribution from diagram 2 is negative, a spin-polarized state of a 1D gas may be energetically more favorable than an unpolarized state.

We limit our discussion to the first two diagrams, which means that the particle exchange interaction is taken into account within the Hartree–Fock approximation. Thus,

$$E_1 = \frac{1}{2} \sum_{E_K, E_L < E_F} \left[ \langle KL|V|KL \rangle - \langle KL|V|LK \rangle \right] \qquad (11)$$

Here, the first term is the Hartree correction, whereas the second term is the Fock correction to the exchange energy. The summation is carried out both over spatial and spin variables. The first term diverges in the thermodynamic limit ($N \to \infty, \Omega \to \infty, N/\Omega = n = const$). This divergence is compensated however by the term describing the interaction with a positively charged background. Thus, in the first order, the exchange interaction plays a decisive role. Below, we consider its behavior in a 1D system.

*2.1. Exchange interaction in a quasi-1D system.*

The matrix element of the exchange interaction for electrons in a quantum wire has the form

$$\langle KL|V|LK\rangle = \frac{e^2}{\Omega_{1D}^2}\int \frac{e^{-ikz'}e^{-ilz''}e^{ilz'}e^{ikz''}}{\sqrt{|\mathbf{\rho}'-\mathbf{\rho}''|^2+(z'-z'')^2}}|\varphi(\mathbf{\rho}')|^2|\varphi(\mathbf{\rho}'')|^2 d\mathbf{\rho}'d\mathbf{\rho}''dz'dz''=$$
$$= \frac{e^2}{\Omega_{1D}}\int \frac{e^{-ikz}e^{ilz}}{\sqrt{|\mathbf{\rho}'-\mathbf{\rho}''|^2+z^2}}|\varphi(\mathbf{\rho}')|^2|\varphi(\mathbf{\rho}'')|^2 d\mathbf{\rho}'d\mathbf{\rho}''dz \qquad (12)$$

where the $z$ coordinate coincides with the wire axis. Here, the transformation of the variables $z=z'-z''$, $Z=(z'+z'')/2$ was performed with the integral over $Z$ being equal to the sample length $\Omega_{1D}$. Thus, the expression for the exchange interaction energy can be written as

$$E_{exc} = -\frac{1}{2}\sum_{K,L<k_f}\langle KL|V|LK\rangle = -g_s\frac{e^2\Omega_{2D}}{2(2\pi)^2}\int_{-k_F}^{k_F}e^{ikz}dk\int_{-k_F}^{k_F}e^{ilz}dl\int \frac{|\varphi(\mathbf{\rho}')|^2|\varphi(\mathbf{\rho}'')|^2}{\sqrt{|\mathbf{\rho}'-\mathbf{\rho}''|^2+z^2}}d\mathbf{\rho}'d\mathbf{\rho}''dz =$$
$$= -g_s\frac{e^2\Omega_{1D}}{2\pi^2}\int \frac{\sin^2(k_F z)}{z^2\sqrt{|\mathbf{\rho}'-\mathbf{\rho}''|^2+z^2}}dz\int |\varphi(\mathbf{\rho}')|^2|\varphi(\mathbf{\rho}'')|^2 d\mathbf{\rho}'d\mathbf{\rho}'' \qquad (13)$$

Subsequent substitution, $u=z/|\mathbf{\rho}'-\mathbf{\rho}''|$ and $\alpha=k_F|\mathbf{\rho}'-\mathbf{\rho}''|$, in the integral over $z$ results in the following expression for the density of the exchange interaction energy in a quasi-1D system

$$\varepsilon_{exc} = E_{exc}/\Omega_{1D} = -g_s\frac{e^2}{2\pi^2}\int \frac{|\varphi(\mathbf{\rho}')|^2|\varphi(\mathbf{\rho}'')|^2}{|\mathbf{\rho}'-\mathbf{\rho}''|^2}I(\alpha)d\mathbf{\rho}'d\mathbf{\rho}'' \qquad (14)$$

Here,

$$I(\alpha) = \int_{-\infty}^{+\infty}\frac{\sin^2(\alpha u)}{u^2\sqrt{1+u^2}}du$$

In the limit of the low linear concentration of the charge carriers

$$\alpha = k_F|\mathbf{\rho}'-\mathbf{\rho}''| \ll 1 \qquad (15)$$

this integral is estimated as

$$I(\alpha) \approx \alpha^2\left(-\frac{1}{2}\ln\alpha + \frac{3}{4} - \frac{C}{2}\right) \qquad (16)$$

Here, $C$ is the Euler constant ($C \approx 0.5772$). Thus, we obtain the following expression for the density of the exchange interaction energy as a function of the concentration of the charge carriers in a 1D system:

$$\varepsilon_{exc} \approx -g_s \frac{e^2 k_F^2}{2\pi^2} \int |\varphi(\mathbf{\rho}')|^2 |\varphi(\mathbf{\rho}'')|^2 \left[ -\frac{1}{2}\ln(k_F|\mathbf{\rho}'-\mathbf{\rho}''|) + \frac{3}{4} - \frac{C}{2} \right] d\mathbf{\rho}' d\mathbf{\rho}'' \approx -\frac{\beta_{1D}}{g_s} n_{1D}^2 + \frac{\gamma_{1D}}{g_s} n_{1D}^2 \ln\left(\frac{n_{1D} R}{\pi g_s}\right)$$

(17)

Here,

$$\beta_{1D} = e^2 \left(\frac{3}{8} - \frac{C}{4}\right) \approx 0.28 e^2$$

$$\gamma_{1D} = \frac{e^2}{4}$$

and $R$ is the width of the quantum wire.

Since this expression is obtained in the limit of the low linear concentration of charge carriers, when

$$k_F R \ll 1, \tag{18}$$

the logarithmic factor in the second term should be noted to be negative that results in the corresponding negative exchange energy correction.

The opposite limiting case when

$$k_F R \sim \alpha \gg 1 \tag{19}$$

is also followed to be analyzed. The integral $I(\alpha)$ is estimated in frameworks of this limit as

$$I(\alpha) = \int_{-\infty}^{+\infty} \frac{\sin^2(\alpha u)}{u^2 \sqrt{1+u^2}} du \approx A\alpha$$

$$A = \int_{-\infty}^{+\infty} \frac{\sin^2(t)}{t^2} dt \approx 3.1375$$

(20)

Then, the linear density of the exchange energy

$$\varepsilon_{exc} = E_{exc}/\Omega_{1D} = -\frac{g_s e^2}{2\pi^2} \int \frac{|\varphi(\mathbf{\rho}')|^2 |\varphi(\mathbf{\rho}'')|^2}{|\mathbf{\rho}'-\mathbf{\rho}''|^2} I(\alpha) d\mathbf{\rho}' d\mathbf{\rho}'' \approx \frac{Ae^2}{2\pi^2} k_F = \chi_{1D} n_{1D}$$

$$\chi_{1D} \approx \frac{e^2}{2\pi} \int \frac{|\varphi(\mathbf{\rho}')|^2 |\varphi(\mathbf{\rho}'')|^2}{|\mathbf{\rho}'-\mathbf{\rho}''|^2} d\mathbf{\rho}' d\mathbf{\rho}''$$

(21)

is independent of the spin factor.

## *2.2. Spontaneous spin polarization due to the exchange interaction energy exceeding the kinetic energy.*

To answer the question of whether the exchange interaction may result in the appearance of a spontaneous spin polarization, we have to compare the total energies of spin-polarized and unpolarized states of a quasi-1D gas of charge carriers. In the limit of the low charge carrier concentration (see (8) and (17)), the energy density of a quasi-1D gas equals

$$\varepsilon^{1D} = \varepsilon_{kin} + \varepsilon_{exc}$$

$$\varepsilon^{1D} = \frac{\pi^2 \hbar^2 n_{1D}^3}{12 m g_s^2} - \frac{n_{1D}^2}{g_s}\left[\beta_{1D} - \gamma_{1D} \ln\left(\frac{n_{1D} R}{\pi g_s}\right)\right] \tag{22}$$

Here, the first and the second terms correspond to the kinetic and the exchange interaction energy, respectively. Thus, the energies of the spin-polarized ($g_s = 1$) and unpolarized ($g_s = 2$) states are equal to

$$\varepsilon^{1D}\bigg|_{g_s=1} = \frac{\pi^2 \hbar^2 n_{1D}^3}{12 m} - n_{1D}^2\left[\beta_{1D} - \gamma_{1D} \ln\left(\frac{n_{1D} R}{\pi}\right)\right]$$

and

$$\varepsilon^{1D}\bigg|_{g_s=2} = \frac{\pi^2 \hbar^2 n_{1D}^3}{48 m} - \frac{n_{1D}^2}{2}\left[\beta_{1D} - \gamma_{1D} \ln\left(\frac{n_{1D} R}{2\pi}\right)\right]$$

If the linear concentration $n_{1D}$ is less than a critical value $n_0$ that results from the equation

$$\varepsilon\bigg|_{g_s=1} = \varepsilon\bigg|_{g_s=2}$$
$$\frac{3\pi^2 \hbar^2 n_0}{24 m} = \beta_{1D} - \gamma_{1D} \ln\left(\frac{2 n_0 R}{\pi}\right), \tag{23}$$

the energy of the exchange interaction exceeds the kinetic energy and, thus, the spin-polarized state is energetically more favorable than the unpolarized state. At the same time, if the linear concentration of the charge carriers exceeds the critical value, $n_0$, and the kinetic energy is dominant, the unpolarized state is energetically more favorable. The value of $n_0$ should be noted to be dependent only on the width of the quantum wire and the effective mass that appears to be a function of the concentration of electrons [24, 25] and holes [26-32] in low-dimensional systems.

Two features of the discussed mechanism of a spontaneous spin polarization in low-dimensional semiconductor systems have to be taken into account. First, the results obtained in the limit of low linear concentration of charge carriers appear to independent of the quantum confinement wave functions that are were explicitly introduced into the consideration of quasi-1D systems. Moreover, the conditions for the appearance of ferromagnetic ordering in quasi-1D systems in the limit of low linear concentration of charge carriers are actually reduced to those obtained for strictly 1D systems, in frameworks of the inclusion of correlation corrections does not destroy the stability of the ferromagnetic state due to the exchange interaction [17–19]. Second, the correlation energy is also taken into account to determine the linear concentration value corresponding to the onset of the Wigner crystallization that competes strongly with the transition to a spontaneously spin-polarized state with extended wave functions [20, 33-36]. However, the Wigner crystallization should take place for $r_s \geq 39$ [35], while a spontaneous spin polarization appears at $r_s = 3.3$ [20], where $r_s$ is the ratio of the potential to the kinetic energy. Thus, the transition to the crystalline state in quasi-1D systems occurs at the concentration of carriers that is two to four orders of magnitude lower than the value corresponding to the transition to a spontaneously spin-polarized state with extended wave functions. The spontaneous spin-polarized state with extended wave functions seems to be expected in a quantum wire at higher values of $n_{2D}$ than in a 2D gas of charge carriers because of an additional partial decay of the kinetic energy with a reduction in the system dimensionality [20].

We stress once more that this consideration corresponds to the limit of low linear concentration of charge carriers, $k_F R \ll 1$, for quasi-1D systems. This circumstance imposes serious restrictions on the width of quantum wires. If this condition is not satisfied, the system should be considered in the limiting case of high charge-carrier concentration, $k_F R \gg 1$, for

quasi-1D systems when an unpolarized state is always energetically more favorable than a spin-polarized state, because the exchange energy becomes to be independent of $g_s$.

**3. Spin depolarization and quenching of the "$0.7 \cdot (2e^2/h)$" feature in the quantum conductance staircase of a quantum wire.**

*3.1. Spin depolarization of electrons in the GaAs based quantum wires.*

The critical linear density $n_0$ corresponding to a complete spin depolarization in a quantum wire connecting two 2D reservoirs, which is given by (23), depends upon both the width of the quantum wire and the effective mass that increases as the value of $n_{2D}$ decreases [24-26]. Such behavior of the effective mass for electrons specifically in the GaAs based quantum wires seems to be caused by its energy dependence that was calculated in the case when the kinetic energy and the quantum-confinement energy are dominant in low-dimensional semiconductor systems [37, 38]:

$$m = m_0(1 + 1.447E + 0.245E^2) \qquad (24)$$

Here, $E$ is the sum of the kinetic and quantum-confinement energies and the coefficients of $E$ account for the band parameters of GaAs.

The electron effective mass value has been found to increase by a factor of $1.1 \div 1.2$ that is due to a rise in quantum-confinement energy as the width of the quantum well decreases below 10 nm [37–39]. The electron effective mass is of importance to rise as the value of $n_{2D}$ increases in the GaAs based quantum wells, even though the exchange energy that compensates the kinetic energy is taken into account in (24). However, the exchange interaction may significantly affect the effective mass of charge carriers in quantum wires, because the spin-polarized states with extended wave functions in a quasi-1D system are spontaneously formed at higher values of $n_{2D}$ or $p_{2D}$ than in a 2D gas [20]. Therefore, the kinetic energy is effectively quenched in the middle part of a quantum wire that connects two 2D reservoirs, because a competition with the exchange energy is available, which may favor a reduction in the effective mass with increasing the values of $n_{2D}$ or $p_{2D}$ (see (22) and (24)). Such a gain in the exchange interaction may account for a rise in the effective mass of electrons as their concentration decreases in quantum wells [25], because the low density 2D gas is able to decay in the system of two-dimensional lakes connected by quantum wires or quantum point contacts, which result from specifically in the presence of disorder [40].

The dependences of the electron effective mass in the GaAs based quantum wells on the value of $n_{2D}$ that were calculated when the exchange energy has been taken into account in the relationship (24) allowed to determine the values of the critical linear concentration, $n_0$ (see (23)), which corresponds to a complete spin depolarization of electrons in the quantum wire connecting two 2D GaAs reservoirs (Fig. 1). Here, these dependences of $n_0$ on the value of $n_{2D}$ are used in analysis of the "$0.7 \cdot (2e^2/h)$" feature in the quantum conductance staircase as a function of the electron concentration in the quantum wire prepared in the GaAs based quantum well by the split-gate method, with the electron sheet density tuned controllably over one order of magnitude by biasing an overall top gate [1] (Fig. 1).

The "$0.7 \cdot (2e^2/h)$" feature is seen to attain almost the value of $0.5 \cdot (2e^2/h)$ at sufficiently small values of $n_{2D}$. Thus, spin degeneracy of the substep in the quantum conductance staircase is lifted, when the 1D channel is completely spin-polarized. However, if the electron concentration in the 2D reservoir attains the value corresponding to the critical linear concentration in the 1D channel, $n_0$, the "$0.7 \cdot (2e^2/h)$" feature evolves towards its normal value because of partial spin depolarization. Besides, the apparent level-off of the "$0.7 \cdot (2e^2/h)$" feature near the value of $0.75 \cdot (2e^2/h)$ appears to be due to its temperature dependence, which results from because of partial spin depolarization of the electron gas near the bottom of the 1D subband at finite temperatures [23].

A most interesting result is the unexpected transformation of the "$0.7 \cdot (2e^2/h)$" feature to the value of $0.5 \cdot (2e^2/h)$ with a subsequent increase in the electron sheet density (see Fig. 1).

This recreation of the $0.5 \cdot (2e^2/h)$ value seems to be caused by also the spin polarization in the quantum wire, which originates probably from the lowest 1D subband that is magnetically ordered by the indirect exchange via electrons excited to the upper subband at a finite temperature. Indirect-exchange mechanisms that cause such non-equilibrium spin polarization in a 1D channel are most probably related to the processes of spin-correlated transport within a narrow band [41, 42] and spin polarization due to the formation of spin polarons [35]. Finally, the $0.5 \cdot (2e^2/h)$ substep is of importance to be observable readily in the quantum wires with a higher level of disorder [43] that is evidence of the indirect exchange in lifting the spin degeneracy of the "$0.7 \cdot (2e^2/h)$" feature in the quantum conductance staircase at large values of $n_{2D}$.

### *3.1. Spin depolarization of holes in the Si based quantum wires.*

` Studies of the quantum conductance staircase revealed by ballistic channels have shown that the "$0.7 \cdot (2e^2/h)$" feature is observed not only in various types of the electron GaAs based quantum wires [7-13, 44-46], but also in the hole Si based quantum wires [47-49]. The latter findings were made possible by the developments of the diffusion nanotechnology that allows to fabricate the ultra-narrow silicon quantum wells of the p-type on the n-type Si(100) surface, which are located between the δ - barriers heavily doped with boron [47, 50, 51].

The angular dependencies of the cyclotron resonance spectra and the conductivity have demonstrated that the self-assembled silicon quantum well (SQW) prepared by short-time diffusion of boron contain the high mobility 2D hole gas with long transport relaxation time of heavy and light holes at 3.8 K, $\tau \geq 5 \cdot 10^{-10} c$ [52-56]. Thus, the transport relaxation time of holes in SQW appeared to be longer than in the best MOS structures [2], contrary to what might be expected from strong scattering by the δ - barriers. This passive role of the δ - barriers between which the SQW is formed was quite surprising, when one takes into account the level of their boron doping, $\approx 10^{21} cm^{-3}$ [47, 51]. To eliminate this contradiction, the temperature dependencies of the conductivity and the Seebeck coefficient as well as the EPR spectra and the local tunneling current-voltage characteristics have been studied [47, 50-55]. The δ - barriers heavily doped with boron appeared to consist of the trigonal dipole centres that seem to result from the negative-U reaction, $2B^0 \rightarrow B^- + B^+$, which define their ferroelectric properties responsible for the suppression of backscattering in the SQW [50-55]. Therefore even with small drain-source voltage the electrostatic ordering within the ferroelectric δ - barriers is able to stabilize the formation of the one-dimensional subbands, when the quantum wires are created inside SQW using the split-gate technique. Thus, the preparation of the narrow p-type SQW confined by the δ - barriers with ferroelectric properties that quench even short-range scattering potential made it possible for the first time to use the split-gate constriction to study the quantum conductance staircase of holes at the temperature of 77 K [47-49, 56].

The device applied here to analyze the dependence of the "$0.7(2e^2/h)$" feature on the concentration of holes in the Si based quantum wires has been firstly advanced in Ref. [1] (see Fig. 2). The basis of this sample is the self-assembled silicon quantum well (SQW) of the p-type that was formed between δ-barriers by the short-time diffusion of boron from the gas phase into the n-type Si (100) surface. The parameters of the SQW that contains the high-mobility 2D hole gas were defined by the SIMS, STM, cyclotron resonance and EPR methods. The sheet density of holes was found to tune controllably over one order of magnitude, from $5 \cdot 10^{12} m^{-2}$ to $9 \cdot 10^{13} m^{-2}$, by biasing the top gate above a layer of insulator, which fulfils the application of the $p^+ - n$ bias voltage. The variations in the mobility measured at 3.8 K that correspond to this range of the values of $p_{2D}$ appeared to occur between 80 and 420 $m^2/Vs$. Thus, the mobility of holes remains high even at low densities. Besides, the high value of mobility appeared to decrease no more than two times in range of temperatures from 3.8 K to 77 K that seems to be caused by both the ferroelectric properties for the δ-barriers and the electric field of the $p^+ - n$ junction [47, 50, 54, 55]. These parameters of 2D hole gas allowed to study the quantum staircase revealed by the heavy holes at 77 K. The experiments were provided by the effective 1D channel length, 0.2 μm, and the QW cross section, 2 nm × 2 nm, which is determined by the SQW width and the lateral confinement due to ferroelectric properties for the δ-barriers. The number of the highest occupied mode of the short quantum wire inserted in the right side arm was controlled by varying the split-gate voltage ($U_g$) (see. Fig. 2).

Figure 3 shows the quantum staircase revealed by the heavy holes in the 1D channel defined by the split-gate voltage inside the SQW provided that the $p^+ - n$ bias voltage is kept to be zero, which appeared to result in the value of $p_{2D}$ equal to $4 \cdot 10^{13} m^{-2}$. Under these conditions, the "$0.7 \cdot (2e^2/h)$" feature is seen to be coincident practically with its normal value. Tuning the value of $p_{2D}$ by biasing the top gate causes however the variations in the height of this feature (Fig. 4). The value of $0.5 \cdot (2e^2/h)$ is found under forward $p^+ - n$ bias voltage, whereas at large values of $p_{2D}$ induced by reverse $p^+ - n$ bias voltage the "$0.7 \cdot (2e^2/h)$" feature attains the value of $0.75 \cdot (2e^2/h)$. The height of the "$0.7 \cdot (2e^2/h)$" feature studied as a function of the hole concentration in the p-type silicon SQW is worthwhile to be related to the behavior of the critical linear concentration, $p_0$, (see (23)), which was calculated by extrapolation from the known dependence of the hole effective mass in the p-type silicon quantum wells on the value of $p_{2D}$ [27-32], with the exchange energy that compensates the kinetic energy (Fig. 5). Thus, the variations in the height of the "$0.7 \cdot (2e^2/h)$" feature that result from controllable tuning the value of $p_{2D}$ can be perceived as a result of partial spin depolarization of holes, which is enhanced as the critical linear concentration in the 1D channel, $p_0$, is approached.

Finally, the behavior of the "$0.7 \cdot (2e^2/h)$" feature in the quantum conductance staircase shares a common trait related to the critical linear concentration of holes and electrons that corresponds to their complete spin depolarization in the 1D channels prepared respectively inside the p-type Si based quantum well studied in this work and inside the n-type GaAs based quantum well discussed above [1].

**4. Conclusions.**

Analysis of the conditions for the appearance of a spontaneous spin polarization in one-dimensional systems placed in a zero magnetic field, which has been carried out within the Hartree–Fock approximation, enabled to determine the critical concentration of carriers that defines a complete spin depolarization of a quasi-1D gas. The range of the linear concentration of carriers that imposes the restrictions to use the plane waves in the studies of the ferromagnetic ordering in one-dimensional systems has been evaluated in the case of dominance of the exchange energy over the kinetic energy.

The transition of a of a quasi-1D gas to the crystalline state has been demonstrated to occur at the concentration of carriers that is two to four orders of magnitude lower than those corresponding to the transition to a spontaneously spin-polarized state with extended wave functions. The spontaneous spin-polarized state with extended wave functions seems to be expected in a quantum wire at higher values of $n_{2D}$ than in a 2D gas of charge carriers because of an additional partial decay of the kinetic energy with a reduction in the system dimensionality.

The dependence of the critical linear concentration that defines a complete spin depolarization in a 1D channel connecting two 2D reservoirs on the carrier's concentration has been derived to analyze corresponding the evolution of the "$0.7 \cdot (2e^2/h)$" feature from the $e^2/h$ to $3/2 \cdot (e^2/h)$ values in the quantum conductance staircase of the quantum wire prepared inside the p-type silicon quantum well using the split-gate technique. The 1D channel studied seems to be spin-polarized at the linear concentration of holes lower than the critical linear concentration, because the "$0.7 \cdot (2e^2/h)$" feature is close to the value of $0.5 \cdot (2e^2/h)$ that indicates the spin degeneracy lifting for the first step of the quantum conductance staircase.

The "$0.7 \cdot (2e^2/h)$" feature has been found, however, to tend to the value of $0.75 \cdot (2e^2/h)$ when the linear concentration of holes attains the critical value corresponding to the spin depolarization.

The variations in the height of the "$0.7 \cdot (2e^2/h)$" feature observed in the hole quantum conductance staircase of the p-type silicon quantum wire seem to be related to the evidences of the quantum conductance staircase obtained by varying the concentration of electrons in the 1D channel prepared inside the GaAs-AlGaAs heterojunction [1].

**ACKNOWLEDGMENTS**

We are grateful to V.F. Sapega and W. Gehlhoff for useful discussions of the results obtained, to V.V. Shnitov for his help in carrying out numerical calculations, and to A.M. Malyarenko for technological assistance.

**Captions**

Fig. 1. Dependence of the critical linear concentration corresponding to a complete spin depolarization of the quasi-1D electron gas in a quantum wire connecting 2D reservoirs in a GaAs/GaAlAs QW on the concentration of electrons; $R = 100 nm, d = 20 nm$. Circles indicate the height of the "$0.7 \cdot (2e^2/h)$" feature determined in the studies of the 1D channels prepared by split-gate method in GaAs/GaAlAs QWs [1, 7-9].

Fig. 2. Schematic diagram of the device that demonstrates a perspective view of the p-type silicon quantum well located between the $\delta$ - barriers heavily doped with boron on the n-type Si(100) substrate as well as the top gate and the depletion regions created by split-gate method, which indicate a D channel connecting two 2D reservoirs.

Fig. 3. The quantum conductance staircase at zero magnetic field that is revealed by the holes in the quantum wire prepared by the split-gate method inside the p-type silicon quantum well on the n-type Si(100) surface. The $p^+ - n$ bias voltage that is fulfilled by biasing the top gate was kept to be zero.

Fig. 4. The quantum conductance staircase of the silicon quantum wire as a function of the sheet density of holes that was tuned controllably by biasing the top gate, which fulfils the application of the $p^+ - n$ bias voltage to the p-type silicon quantum well on the n-type Si(100) surface. The $p^+ - n$ bias voltage is varied from the forward branch to the reverse branch between $+110 mV$ and $-120 mV$, which establish the range of the magnitude of $p_{2D}$ from $5 \cdot 10^{12} m^{-2}$ to $9 \cdot 10^{13} m^{-2}$ provided that zero $p^+ - n$ bias voltage results in the value of $p_{2D}$ equal to $4 \cdot 10^{13} m^{-2}$.

Fig. 5. Dependence of the critical linear concentration corresponding to a complete spin depolarization of the quasi-1D electron gas in a quantum wire connecting 2D reservoirs in a silicon quantum well on the concentration of holes; $R = 2 nm, d = 2 nm$. Circles indicate the values of the height of the "$0.7 \cdot (2e^2/h)$" feature that are shown in Fig. 4.

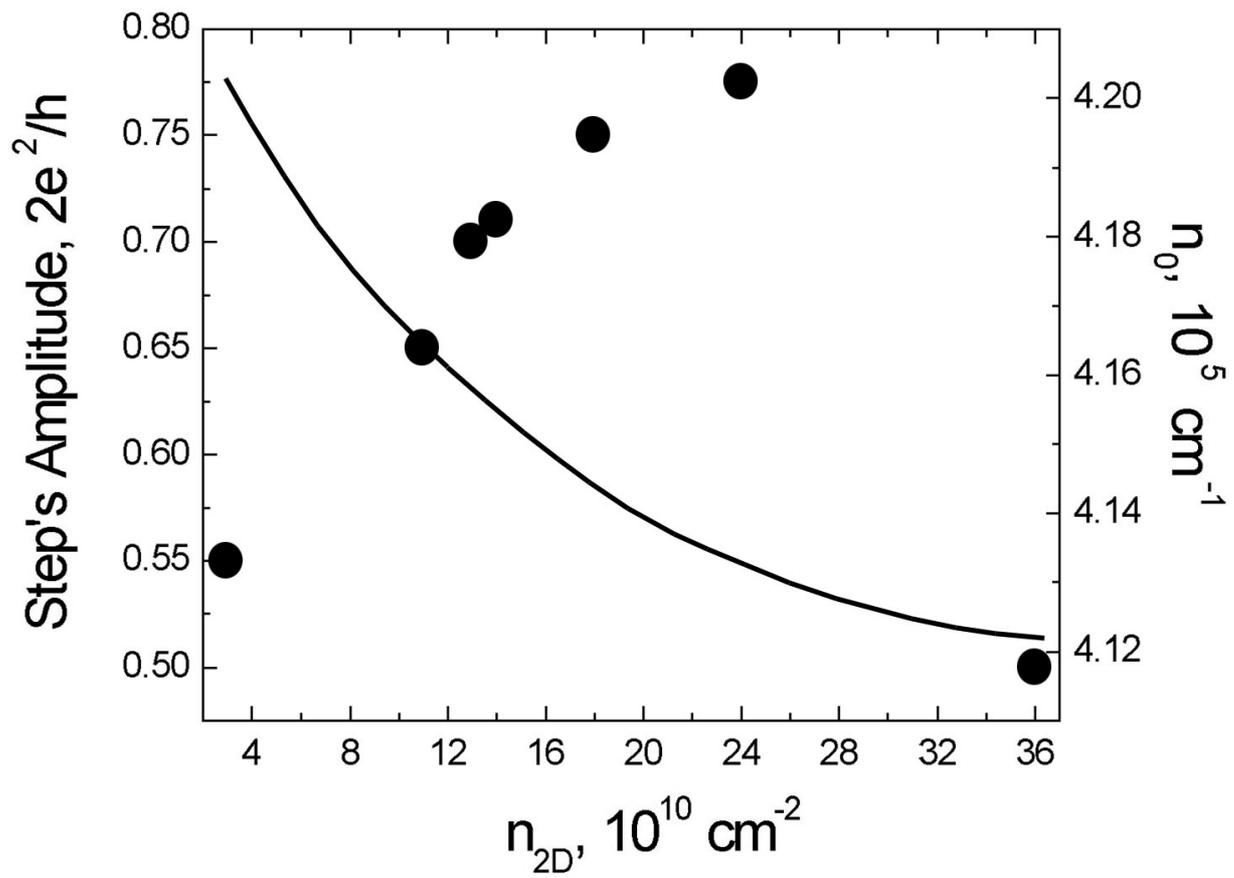

Fig.1.

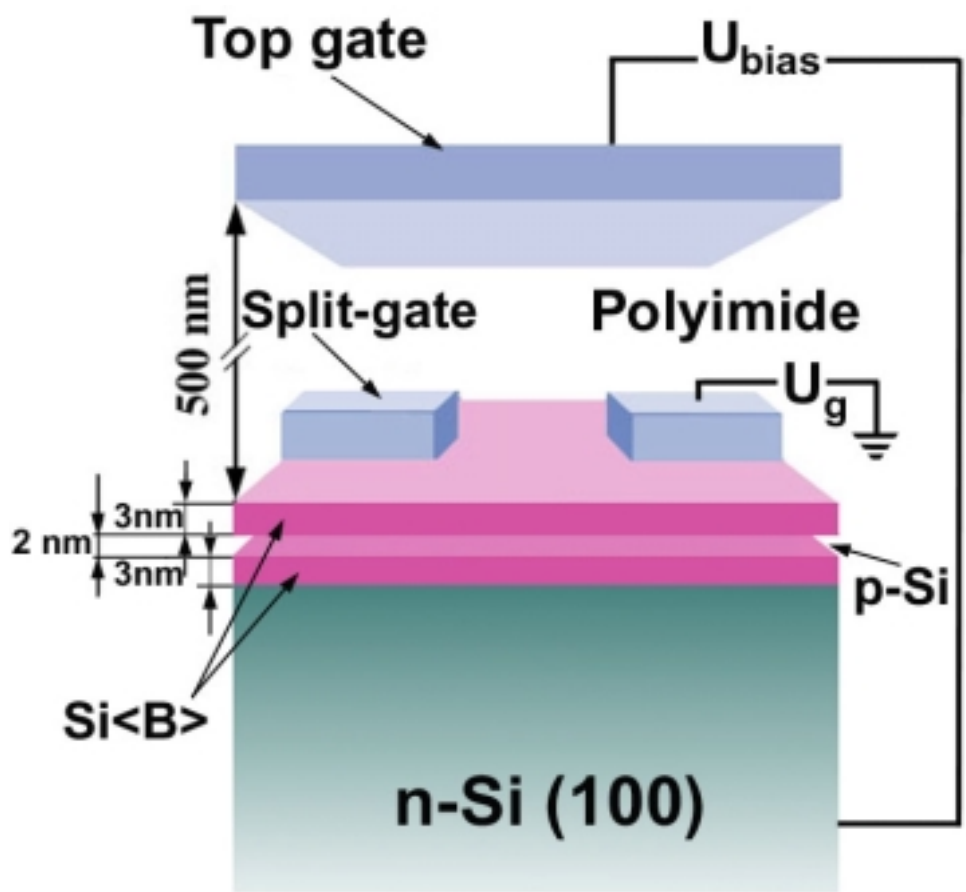

Fig.2.

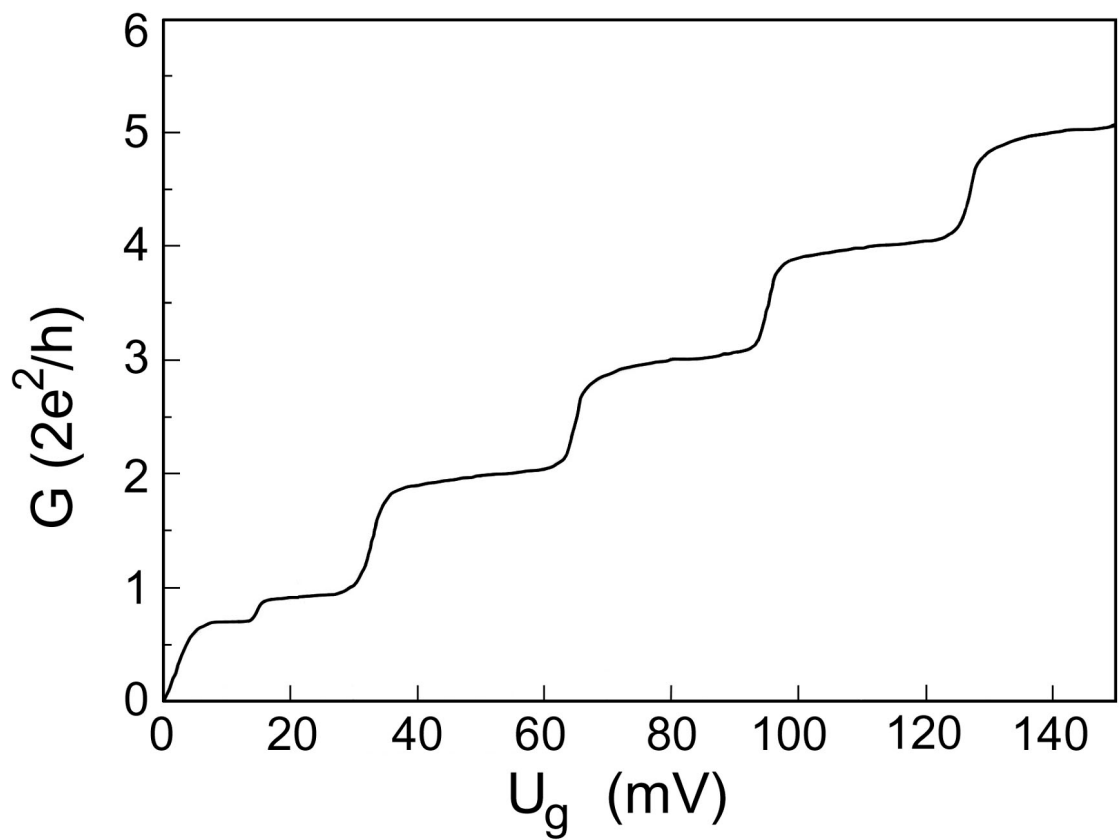

Fig.3.

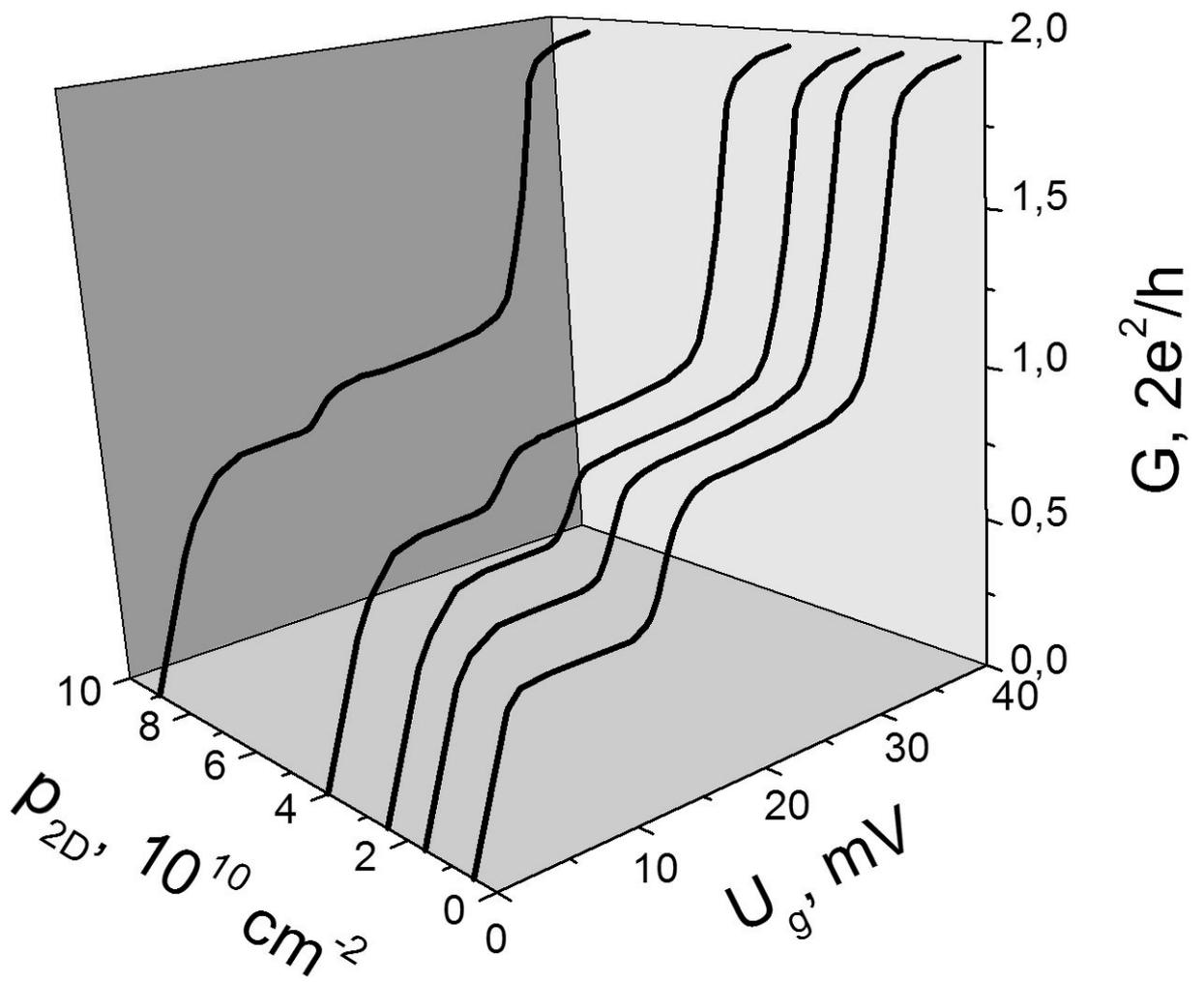

Fig.4.

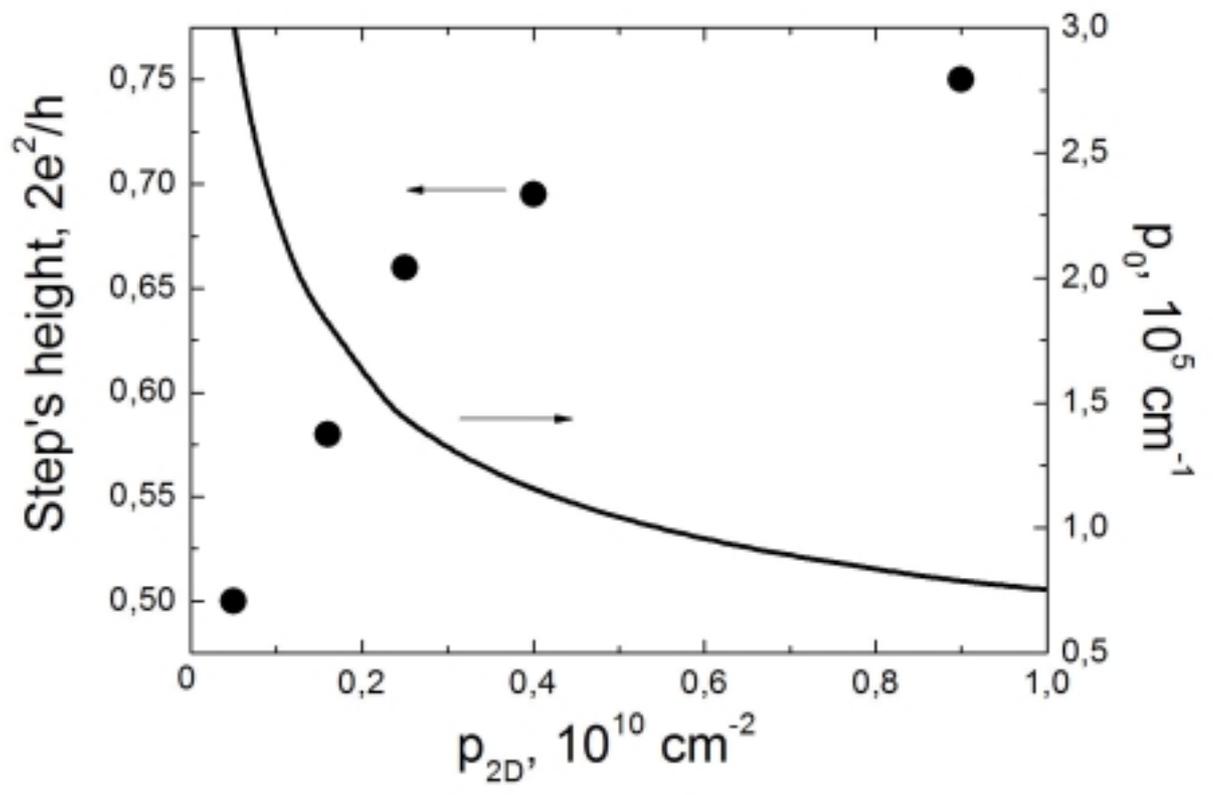

Fig.5.